---------------------------------------------------------------

\documentstyle[prd,aps,preprint]{revtex}

\let\ssection=\section
\renewcommand{\section}{\setcounter{equation}{0}\ssection}
\def\sqr#1#2{{\vcenter{\hrule height.#2pt\hbox{\vrule width.#2pt
-1zheight#1pt \kern#1pt \vrule width.#2pt}\hrule height.#2pt}}}

\newcommand{\be}{\begin{equation}}
\newcommand{\ee}{\end{equation}}
\newcommand{\ben}{\begin{eqnarray}}
\newcommand{\een}{\end{eqnarray}}
\newcommand{\bec}{\begin{center}}
\newcommand{\eec}{\end{center}}

\begin{document}
\preprint{gr-qc/9407041}

\draft
\widetext

\title{A new method of generating exact inflationary solutions}

\author{Franz E.~Schunck\footnote{Electronic address:
fs@thp.uni-koeln.de} and Eckehard W.~Mielke\footnote{Electronic address:
pke27@rz.uni-kiel.d400.de}}

\address{Institute for Theoretical Physics, University of Cologne,
D--50923 K\"oln, Germany}

\date{\today}

\maketitle

\begin{abstract}
The mechanism of the initial inflation of the universe is based on
gravitationally coupled scalar fields $\phi$. Various scenarios are
distinguished by the choice of an {\it effective self--interaction
potential} $U(\phi)$ which simulates a {\it temporarily}
non--vanishing {\em cosmological term}. Using the Hubble expansion
parameter $H$ as a new ``time"
coordinate, we can formally derive the {\it general} Robertson--Walker
metric for a {\em spatially flat} cosmos. Our new method provides a
classification of allowed inflationary potentials and is broad enough to
embody all known {\it exact} solutions involving one
scalar field as special cases.
Moreover, we present new inflationary and deflationary exact solutions
and can easily predict the influence of the form of $U(\phi)$ on
density perturbations.
\end{abstract}
\bigskip\bigskip

\narrowtext

\section{Introduction: Einstein's biggest blunder}

The introduction of a {\em cosmological constant} ${\Lambda }$ in the field
equations of general relativity later on stroke
Einstein as ``the biggest blunder of my life'' \cite{1,mtw}.
Such an amendment was not even completely new, since von Seeliger
\cite{3} and Neumann \cite{4}, e.g., have considered
already in 1896 a corresponding term in the Poisson equation for the
Newtonian potential in
order to compensate the energy density of the `\ae ther'.

Nowadays, Einstein's dream of a completely geometrical description
of fundamental physical interactions has evolved into supergravity \cite{5}
and superstring models \cite{6} in a way which was unprecedented at his
times.
Nevertheless, the cosmological term is still a major problem of these
new approaches, as can be inferred from the review of Weinberg \cite{7}.

The overall reason being that, in almost all quantized theories of
particle interactions, the vacuum density
$\rho_{vac}$ gives rise to
a huge {\em bare} cosmological constant $\Lambda_0 =
\kappa \rho_{vac}$. This can be traced back
to the fact that the vacuum
fluctuations feel all the complicated physics originating from
Higgs fields, fermion
condensates etc., which enter into today's unified field theories.
For much higher energies or, equivalently, to very short spacetime
distances, the small scale behavior of the quantum world
would determine the large scale structure of the universe.

On the other hand, it is known that the observed macroscopical energy density
${\epsilon}$ is extremely small. For the range of
45--100 km s$^{-1}$ Mpc$^{-1}$ of {\em today's}
Hubble constant $H_0$, the critical density is estimated as
$\rho_c = 0.5 - 2 \times 10^{-29}$ g/cm$^3$. From
local as well as large scale astronomical measurements, respectively,
the macroscopically observed cosmological constant
${\Lambda}$ is estimated \cite{8} to be less than
$4\times 10^{-56}$ cm$^{-2}$.
Since the vacuum energy may also be
time--dependent at the early stages of the universe, the exact
fine--tuning of the various vacuum contributions to a very small
${\Lambda}$ in the low
temperature regime of today appears to be one of the great
mysteries about unification.

Higgs--type scalar fields become more and more important. They are not only
induce the masses for the elementary particles via the Higgs--Kibble
mechanism, but they can also form
stable boson stars \cite{kusm} and kinks \cite{baek}. For
spin--one--particles, exact non--singular solutions of the
Einstein--$SU(2)$--Yang--Mills system are not yet known, but the new
power series expansion technique of Ref.~\cite{sch2} can be regarded as a first
attemption in this direction.

But the scalar fields, in disguise as the ``inflaton'' $\phi $, can
also dominate the early universe, the {\em epoch of inflation}.
Before symmetry breaking, a self--interaction $U(\phi )$ of such
gravitationally coupled scalar fields allows us to introduce a
{\em variable} ``cosmological term" without violating the
Noether--Bianchi identities of Einstein's general relativity.

\section
{\bf Model of a universe with inflation}

$\!$
 From new astronomical observations (COBE) we know that the universe expands,
is rather homogenious on
the large scale and in the microwave background.
However, the standard Friedmann model of the cosmos offers no solution
to such issues as the
singularity problem, the problem of flat space, the horizont problem,
the homogenity problem on great scales, the absence of magnetic
monopoles \cite{mie86}, and the problem of large number of particles
\cite{linde,brandenberger}.

The idea of {\em inflation} (see Guth \cite{guth} and Linde \cite{linde82})
attempts to solve several of these problems.
Scalar fields (Higgs, axion) are expected to generate, shortly after the
big bang, an exponential increase of the universe. However, in
these first attempts, there was no so--called {\em graceful
exit} to the Friedmann cosmos, and the inflationary phase did not end.
This problem was
solved in the {\em new inflationary universe}. In this model, the
scalar field is ruled by a slightly different self--interaction
potential which possesses a slow--roll part (a plateau)
of the potential (acting as a vacuum energy) which
dominates the universe at the beginning.

Later on,
power--law models were constructed which possess no
exponential but an $a(t) \sim t^n$ increase of the expansion factor of
the universe \cite{lucmat,barrow87}.
The {\em intermediate inflation} is merely
a combination of exponential and power--law increase
\cite{barrow90}. Mathematically, inflation is described by a positive
second derivative of the scale factor $a(t)$ of the universe. In
general, this requires $\rho + 3p < 0$, where $\rho $ is the
density and $p$ the pressure of the matter field.

In the models of the new and chaotic inflationary models, we have
a fine--tuning problem which consists of the combination of the
largeness of the scale factor and an acceptable distribution for density
perturbations \cite{stein}. Solutions for these problems are attempted
in the scenarios of {\em extended inflation} \cite{stein,laste,barmae}.

In all of these models, the isotropy and homogenity are prescribed. It was
also shown that all initially expanding homogeneous models (the Bianchi
and Sachs--Kantowski universes) which
include a positive cosmological constant, approach asymptotically the
de Sitter solution \cite{wald,turner} which is isotropic.
This is called the ``cosmic
no--hair'' theorem. For models with scalar inflation, the question of
damping a possibly initial anisotropy of the universe is not
relevant, because the model merely has to a ensure a very small anisotropy
in the universe after inflation \cite{barrow93}.

In this paper, we present in Sec.~5 a general inflationary solution in
terms of the Hubble parameter
which comprise all previous exact solutions.
This enables us, in Sec.~6, to
classify the potential $U(\phi )$ for the scalar field according to
the different onset of inflationary, deflationary, and Friedmann phases
of the universe.
Within this new description, some new exact
solutions of the so--called {\em new} and {\em chaotic} type are
found in Sec.~8, 9, and 10.
The potentials found have a rather complicated form which, however,
 have so far no motivation from field theory.

Recently {\it chaotic models} with {\em several} scalar
fields $\phi_I$ including the inflaton have attracted much
attention. Linde termed his model ``hybrid inflation"
\cite{lin94}, cf. Copeland et al.~\cite{cope94}.
In the vacuum dominated
regime, the back reaction of the other scalar field on the inflaton
is negligible and we can follow their evolution explicitly by using
quantum field theory in curved spacetime, in the case of a de Sitter
background, see \cite{mie77}, for example.
In these hybrid models, all the other scalar fields vanish, if they sit
in the false vaccum. For the remaining inflaton, we can then simply
apply the general solution of Sect.~\ref{7}.

\section {\bf Friedmann spacetime}

For a rather general class of inflationary models the Lagrange density reads
\be
{\cal L} = \frac{1}{2 \kappa } \sqrt{\mid g \mid}
 \Biggl ( R
   + \kappa \Bigl [ g^{\mu \nu } (\partial_\mu \phi ) (\partial_\nu \phi )
   - 2 U(\phi ) \Bigr ] \Biggr )  \; , \label{lad}
\ee
where $\phi $ is the scalar field and $U(\phi )$ the self--interaction
potential. We use natural units with $c=\hbar =1$.
A constant potential $U_0= \Lambda / \kappa $ would simulate the
cosmological constant $\Lambda $. Scalar coupled Jordan--Brans--Dicke type
\cite{brans}
models can be reduced to (\ref{lad}) via the Wagoner--Bekenstein--Starobinsky
transformation \cite{wag,beken,kasp,galt}. We are looking for
solutions of the Einstein equation
\be
R_{\mu \nu } - \frac {1}{2} g_{\mu \nu } R
  = - \kappa T_{\mu \nu } \; ,
\ee
which are of the Robertson--Walker type
\ben
ds^2 & = & dt^2 - a^2(t) \left [ \frac{dr^2}{1 - k r^2} + r^2 \left (
       d\theta^2 + \sin^2 \theta d \varphi^2 \right ) \right ]
\; , \nonumber \\
& & \qquad k = 0, \pm 1 \; , \label{rw}
\een
where $a(t)$ is the expansion factor with the dimension {\em length}.
An open, flat, or closed universe is characterized by $k=-1,0,1$,
respectively.
This means that we will investigate {\em homogeneous and isotropic} spacetimes.
The scalar field depends only on the time $t$, i.e.~$\phi = \phi(t)$.
Then, the only non--vanishing components of the energy--momentum tensor read
\ben
\rho & = & T_0{}^0 = \frac {1}{2} \dot \phi^2 + U  \; , \\
p & = & - T_1{}^1 = - T_2{}^2 = - T_3{}^3 = \frac {1}{2} \dot \phi^2 - U
 \; .
\een

\section {\bf Reparametrized self--interaction}

Let us assume that $a(t) \neq 0$, furthermore, we express our result
in terms of the Hubble expansion rate
\be
H := \frac {\dot a(t)}{a(t)} \; .  \label {H(t)}
\ee
Only the diagonal components of the Einstein equation are non--vanishing. The
$(0,0)$ component is
\be
3 \left ( H^2 + \frac {k}{a^2} \right ) =
  \kappa \rho   \; . \label{1}
\ee
It describes the conservation of the energy. The $(1,1)$,
$(2,2)$, and $(3,3)$ components are given by
\be
2 \dot H + 3 H^2 + \frac {k}{a^2} = - \kappa p  \; . \label{2}
\ee
The resulting Klein--Gordon equation is
\be
\ddot \phi = - 3 H \dot \phi - U'(\phi )
\; , \label{scalar}
\ee
which, after multiplication by $\dot \phi $, can be transformed into
\be
\frac {1}{2} ((\dot \phi)^2) \dot{} = - 3 H (\dot \phi )^2
 - \dot U \; . \label{scatra}
\ee
 From (\ref{1}) and (\ref{2}) we obtain by linear combination
\be
\dot H = \frac {k}{a^2} - \frac {\kappa }{2} (\rho + p) =
 \frac {k}{a^2} - \frac {\kappa }{2} \dot \phi^2  \label {2'}
\ee
and
\ben
\dot H + 3 H^2 + \frac {2k}{a^2} & = & \frac {\kappa }{2} (\rho - p)
\label {min}\\
 & = & \kappa U  \; . \nonumber
\een
Observe that (\ref{min}) is, in view of (\ref{scatra}) and (\ref{2'}),
a {\it first integral} of (\ref{scalar})
for all values of the normalized extrinsic curvature scalar $k$.
Alternatively, if we eliminate the $k/a^2$ terms in
equations (\ref{1}) and (\ref{2}), we obtain the Raychaudhuri equation
\be
\dot H + H^2 = \frac {\ddot a}{a} = - \frac {\kappa }{6} (\rho + 3p) \; .
\label {ray}
\ee

There are several options to calculate solutions for the given
system of differential equations (\ref{1}) and (\ref{2}) or
(\ref{2'}) and (\ref{min}), respectively.
The first possibility is to assume a reasonable functional dependence of
the scale factor
$a(t)$ and then to calculate simply the Hubble expansion rate
$H(t)$. However, even for $k=0$ the equation (\ref{2'}) is not easily
integrable in closed form.

Secondly, one could imagine a
potential $U(\phi )$ which possesses the physically desirable
features, and consider (\ref{2'}) and (\ref{min}), which, for $k=0$,
form the autonomous nonlinear system
\ben
\dot H & = & \kappa U(\phi ) - 3H^2 \; , \label{doth} \\
\dot \phi & = & \pm \sqrt {\frac {2}{\kappa }}
  \sqrt{3H^2 - \kappa U(\phi )} \; .\label{dotphi}
\een
In the phase space \cite{piran}, the equilibrium states of this system are
given by the
constraint $\{ \dot H,\dot \phi \}=0$. This constraint is fulfilled by
$\kappa U(\phi ) = 3H^2$, where the Hubble expansion rate is constant,
i.e.~$H_0=\sqrt{\Lambda /3 \,}$. For $\dot \phi=0$, we obtain
a de Sitter--type inflation with $a(t)=\exp (\sqrt{\Lambda /3 \, } t)$.

For $\kappa U(\phi ) \neq 3H^2$, we find
$\{ \dot H, \dot \phi \} \neq 0$, which implies that the solution
$\phi = \phi (t)$ and $H=H(t)$ are {\em invertible}. Then we can write the
potential in (\ref{doth}) and (\ref{dotphi}) in the {\em reparametrized} form
\be
U(\phi ) = U(\phi (t)) = U(\phi (t(H))) = \widetilde U(H)  \; . \label{uh}
\ee
Another question is whether it is possible to construct
$H=H(t)$ from the {\em inverse} function $t=t(H)$ in closed form.
Only in this case, the Hubble expansion parameter and
the scalar field can be expressed explicitly as a function of time,
and the self--interaction potential $U(\phi )$ can be recovered from
$\widetilde U(H)$.

\section {\bf General metric of a spatially flat inflationary universe
\label{7}}

In view of (\ref{1}), (\ref{2}), and (\ref{uh}), for $k=0$, the density and
the pressure can be reexpressed as
\ben
\rho & = & \frac {3}{\kappa } H^2
 \label{dens} \; , \\
p & = & - \rho - \frac {2 \dot H}{\kappa } \\
& = & \rho - 2 \widetilde U  \label{rho-2u} \; .
\een
Hence, the density $\rho $ is always a positive function,
whereas the pressure $p$ is
indefinite and changes sign at $\kappa \widetilde U = 3H^2/2$.

For $\kappa \widetilde U \neq 3H^2$, we find from (\ref{doth}) and (\ref{uh})
the formal solution for the coordinate time
\be
t = t(H) = \int
   \frac {dH}{\kappa \widetilde U - 3 H^2}  \label{tH} \; .
\ee
In formal expressions involving indefinite integrals,
we omit the constant of integration.
The scale factor in the metric follows from the definition
(\ref{H(t)}) of the Hubble expansion rate as $a(t)=a_0 \exp (\int H dt)$
where $a_0$ is a constant with dimension {\em length}.
Inserting (\ref{tH}), we thus can determine the general
solution as
\be
a = a(H) = a_0 \exp \left (
   \int \frac {H dH}{\kappa \widetilde U - 3 H^2} \right )  \label{aH} \; .
\ee

This implies for $k=0$ that the reparametrized Robertson--Walker metric for
inflation reads
\ben
ds^2 & = &
\frac {dH^2}{\left (\kappa \widetilde U - 3 H^2 \right )^2}
 - a_0{}^2 \exp \left ( 2
   \int \frac {H dH}{\kappa \widetilde U - 3 H^2} \right ) \times  \nonumber \\
 & &   \left [ d r^2 + r^2 \left (
    d \theta^2 + \sin^2 \theta d \varphi^2 \right ) \right ] \; .
\een
Note that the Hubble expansion rate $H$ has become the (inverse)
time coordinate. This resembles the reparametrization of
Hughston, cf.~\cite[p.~731]{mtw}, for the Friedmann solution,
in which $a(t)$ serves as the new time
parameter. In view of (\ref{tH}), the general solution of (\ref{dotphi})
for the scalar field can be calculated in terms of the Hubble
parameter $H$ as
\be
\phi = \phi (H) = \mp \sqrt {\frac{2}{\kappa }}
 \int \frac {dH}{\sqrt{3H^2-\kappa \widetilde U}} \; . \label{phiH}
\ee
Our general formula (\ref{phiH}) resembles the Wagoner--Starobinsky
transformation from the conformal Brans--Dicke frame to an Einstein frame,
cf.~\cite{kasp}. In metric--affine gauge
theories of gravity \cite{hehl} this transformation has a rather
natural origin from generalized conformal changes of the metric.

If we introduce the conformal time $T$ via $dt=a(t)\, dT$, the
Robertson--Walker metric (\ref{rw}) (for $k=0$) acquires the manifest
conformally flat form
\be
ds^2= a^{2}(t) \left [ dT^2 - dr^2 - r^2 \left (
       d\theta^2 + \sin^2 \theta d \varphi^2 \right ) \right ] \; .
\ee
For our general solution, the conformal time can be expressed by the relation
\be
dT= {{dH}\over{\kappa \widetilde U -3H^2}}\,
\exp\int {{HdH}\over{3H^2-\kappa \widetilde U}} \; .
\ee

Our general solution holds for $k=0$ and for
$\widetilde U \neq 3H^2/\kappa $.
Since this singular case leads to the de Sitter inflation,
we try in the explicit models the ansatz
\be
\widetilde U(H) = \frac {3}{\kappa } H^2 + \frac {g(H)}{\kappa}
\ee
for the potential, where $g(H)$ is a nonzero function for the
{\em graceful exit}.

\section
{\bf Allowed inflationary potentials}

Inflation and deflation necessarily occur for all
potentials which satisfy the matter condition $\rho + 3p<0$,
i.e.~$\ddot a(t)>0$.
In order to discriminate inflationary from deflationary models,
one has to take into account also the rate of change
of the scale factor $a(t)$ or the sign of the Hubble expansion rate,
respectively. For {\em inflation}, we require
\be
\ddot a > 0 \; , \quad \dot a > 0 \quad \Longleftrightarrow \quad  H>0 \; ,
\ee
whereas for {\em deflation} we require
\be
\ddot a > 0 \; , \quad \dot a < 0 \quad \Longleftrightarrow \quad  H<0 \; .
\ee

For a classification of the potentials, we follow
Ref.~\cite[p.~773]{mtw} and call
\ben
q(t) & := & - \frac {\ddot a a}{\dot a^2}
   = - \left (1 + \frac {\dot H}{H^2} \right ) \\
 & = & 2 - \kappa \frac {\widetilde U}{H^2}
\een
the {\it deceleration parameter}.
Because $\dot a^2$ and $a$ are positive, an accelerating
cosmos $(\ddot a >0)$ is described by negative $q$ values.
Thus, acceleration can only occur for a potential
satisfying
\be
\kappa \widetilde U > 2 H^2 \; .
\ee

According to (\ref{rho-2u}) the pressure is then necessarily negative
and {\em drives} the inflation.
Another constraint is found by looking at the general solution for the
scalar field (\ref{phiH}). The scalar field remains real only if the
potential fulfills $\kappa \widetilde U < 3 H^2$. [Otherwise, we would have a
scalar ``ghost" in the Lagrangian (\ref{lad}).]
 From Fig.~\ref{fig.0}, we can read off the different regions of the
potential. All values within the parabola,
i.e.~$\kappa \widetilde U > 3 H^2$, are
forbidden. All points on the curve $\widetilde U=3 H^2$ are singular for our
system (\ref{doth}) and (\ref{dotphi}) and describe the de Sitter
solution. The origin is the flat and empty Minkowski spacetime.
Solutions within the domain
\be
2H^2< \kappa \widetilde U<3H^2 \quad {\mbox{and}} \quad H>0
\label{infl} \; ,
\ee
bounded by parabolas, describe universes with {\em inflation}.
Solutions within the domain
\be
2H^2< \kappa \widetilde U<3H^2 \quad {\mbox{and}} \quad  H<0 \label{defl}
\ee
describe universes with {\em deflation}.
If these solutions leave this area through $\kappa \widetilde U=2H^2$, they
make contact with a Friedmann cosmos for which $\ddot a <0$.

According to \cite{mie77}, the discrimination between inflation and
deflation depends on the choice of the conformal frame.
For the scalar matter, we find from (\ref{ray}) that
$\rho + 3p=2 (\dot \phi^2 - \widetilde U)$. For
potential--dominated eras this term is negative and hence inflation
can occur. The condition for inflation $\ddot a >0$ is then
equivalently to $\dot H > -H^2$.

For $k=0$ and scalar matter, we can infer from (\ref{2'})
that always $\dot H<0$, i.e.~$-H^2<\dot H < 0$. For other types of
matter it is possible that $\dot H>0$, cf.~eg.~the spin driven
inflation \cite{obuk}. Such physical
models are also called {\em superinflationary}, in contrast to the
{\em subinflationary} ones \cite{kolb,lid,lucmat} considered here.

Several models are now conceivable which have combinations of in- and
deflationary potentials. One can construct models where inflation
never ends, those with a combined inflation--Friedmann cosmos,
or some where the universe enters the deflationary regime. In the
following, we will recover known models from our general formalism and
present some new ones, too.

\section
{\bf Power--law and intermediate inflation}

The ansatz
\be
g(H) = - A \, H^{n} \; ,
\ee
where $n$ is real and $A$ a positive constant, leads to several known
and new solutions. The integration constants $C_1,C_2,C_3$ are, of
course, different in every model. As it turns out, $n=0,1,2$ are
special cases which we consider first:

For $n=0$, we find the following solution:
\be
H(t)=- (A t + C_1)  \; ,
\quad a(t) = a_0 \exp \left ( -\frac{1}{2A} (A t+C_1)^2 + C_2 \right ) \; ,
\ee
and
\be
\phi (t) = \pm \sqrt{ \frac {2A}{\kappa } } (At+C_1-C_3)  \; .
\ee
The chronology in this model is the following: at first, there is an
inflationary phase for which the maximal size of the universe is
$a_0 \exp (-C_1/(2A)+C_2)$. This is very extended for $C_1<<0$ or $C_2>>0$.
The transition from inflation to the standard Friedmann cosmos occurs at the
point $H=+\sqrt{A}$. There exists also a transition from the standard
Friedmann cosmos to deflation
which occurs at $H=-\sqrt{A}$. This all can also be recognized by looking at
the classification diagram (Fig.~\ref{fig.0}).
The self--interaction potential is only
{\em quadratic} as it is normally investigated in the chaotic scenario:
\be
U(\phi ) = \frac {1}{\kappa }
  \left [ 3 \left ( \sqrt { \frac {\kappa }{2A}} \phi + C_3 \right )^2
 - A \right ] \; .
\ee

For $n=1$ we have
\be
H(t)= C_1 \exp (-A t) \; ,
\quad a(t) = a_0 \exp \left (- \frac{C_1}{A} \exp (-At)
 + \frac {C_2}{A} \right ) \; ,
\ee
and
\be
\phi (t) = \pm \sqrt{ \frac {8}{A \kappa } }
 \left [ \sqrt{C_1} \exp \left ( \frac {-At}{2} - C_3 \right ) \right ] \; ,
\ee
so that $C_1>0$. The universe in this model starts with an
inflationary phase up to the point $H=A$, where it crosses the
boundary $\kappa \widetilde U = 2H^2$ and evolves towards
a conventional Friedmann cosmos. The universe reaches the size
$a_0 \exp (C_2/A)$ after infinitely long time.
The potential
\be
U(\phi ) = \frac {A}{8} e^{2 C_3} \phi^2
  \left ( \frac {3 \kappa A}{8} e^{2 C_3} \phi^2
         - A \right )
\ee
has here a linear combination of $\phi^2$-- and $\phi^4$--terms which
are familiar from the Higgs potential of spontaneous symmetry breaking.
For a pure $\phi^4$--potential, an exact and an approximate solution
is found in Refs.~\cite{gott} and \cite{lin83}.

For $n=2$ we find
\be
H(t)=\frac {1}{At+C_1}  \; , \quad  a(t) = a_0 (C_2 (At+C_1))^{1/A}  \; ,
\ee
and
\be
\phi (t) = \pm \sqrt{\frac{2}{A\kappa }}
 \ln \left ( \frac {1}{C_3 (At+C_1)} \right ) \; .
\ee
The self--interaction is the exponential potential
\be
U(\phi ) = \frac {3-A}{\kappa } C_3{}^2
 \exp (\pm \sqrt{2 \kappa A}\; \phi ) \; .
\ee
This case describes power--law inflation $t^{1/A}$ if
$0<A<1$, which means that $2H^2< \kappa U<3H^2$ and $H>0$. For $A=3/2$
the pressure (\ref{rho-2u}) of the scalar field vanishes and we get
$a(t) \simeq t^{2/3}$ as in the matter--dominated Friedmann cosmos
\cite{steph}.
One recognizes that for $A=3$ the scalar field possesses a vanishing
potential (see the Appendix).

For $n\neq 0,1,2$ the constant $A$ has the dimension $length^{n-2}$.
For the Hubble expansion rate, we get
\be
H = \left ( A (n-1) (t+C_1) \right )^{1/(1-n)} \; ,
\ee
whereas the scale factor reads
\be
a(t) = a_0 \exp \left [ (A (n-1))^{1/(1-n)} \frac {1-n}{2-n}
(t+C_1)^{(2-n)/(1-n)} \right ] \; .
\ee
The scalar field is then given by
\be
\phi (t) + C_3 =
\sqrt { \frac {2}{A \kappa } }
\frac {2}{2-n} \Bigl [ A (n-1) (t+C_1) \Bigr ]^{(2-n)/(2(1-n))}  \; .
\ee
The corresponding potential reads
\ben
U(\phi ) & = & \frac {1}{\kappa }
\left [ \sqrt{\frac {\kappa A}{8}} (2-n)
\left ( \phi + C_3 \right )^{2/(2-n)} \right ] \times \nonumber \\
 & &  \left ( 3 \frac {\kappa A}{8} (2-n)^2 \left ( \phi + C_3 \right )^2
 - A \left (\frac {\kappa A}{8} \right )^{n/2} (2-n)^n
 \left ( \phi + C_3 \right )^n
 \right )  \; . \label{un}
\een
Hence, we recover the models \cite{barrow90} of intermediate inflation
$\exp (t^{(2-n)/(1-n)})$ with $0<(2-n)/(1-n)<1$, which is equivalent to
$1<n<2$. For
$(2-n)/(1-n)=2/3$, the flat Harrison--Zel'dovich spectrum is recovered.
Translated into our model this holds for $n=4$, i.e.
\be
\kappa \widetilde U = 3H^2 - AH^4 \; . \label{n=4}
\ee

\section {\bf New potential in the framework of chaotic inflation}

Another model can be obtained from the ansatz
\be
g(H) = \pm \frac {4}{C} \sqrt{A C H - H^2}\, H \; ,
\ee
where $A,C$ are constants. For this model we find the scale factor
\be
a(t) = a_0 \exp [C \arctan (At+B)+F] \; ,
\ee
see Fig.~\ref{fig.1.1}.
 For
\be
a(t) = a_0 \exp [C \mbox{arccot} (-A t)+F] \; ,
\ee
we find the same model, but the scale factor reaches for
$t \rightarrow \infty $ another limit $(A>0)$: $ a_0 \exp ( C \pi )$.

The Hubble expansion rate is
\be
H(t) = \frac {A C}{1+(At+B)^2} \; ,
\ee
which vanishes for $t \rightarrow \infty $.
The solution of the scalar field (Fig.~\ref{fig.1.2}) was determined
via the computer algebra system {\em Macsyma} as
\ben
\phi (t) & = & \pm \sqrt{ \frac {2C}{\kappa }} \Biggl \{
\biggl [ \arctan \Bigl ( \sqrt{2(At+B)} +1 \Bigr )
\nonumber \\
& & + \arctan \Bigl (\sqrt{2(At+B)} -1 \Bigr ) \biggr ] \nonumber \\
& & - \frac {1}{2} \biggl [ \ln \Bigl (At+B+\sqrt{2(At+B)}+1
\Bigr ) \nonumber \\
& & - \ln \Bigl (At+B-\sqrt{2(At+B)}+1 \Bigr ) \biggr ]
  \Biggr \} + C_1  \; . \label{phi1}
\een

In the following, we will only consider this model with positive scalar
field. The potential $U$ depending on the time $t$ reads
\be
U(t) = \frac {A^2 C}{\kappa } \frac {3C - 2At - 2B}{[1+(At+B)^2]^2} \; .
\ee
Because of the complicated functional dependence of $\phi (t)$, we have not
found a closed form of the inverse function, but have numerically
determined $U(\phi )$, see Fig.~\ref{fig.1.3}.
This potential has {\em no} plateau at
the origin $\phi =0$. But it possesses a minimum with a negative
value of $U$. After the minimum a limiting point follows, for which
the potential vanishes: The scalar field needs infinitely long in order to
reach this point.

Eq.~(\ref{phi1}) gives the constraints $C>0$ and $At+B>0$ on the
integration constants. Real solutions occur only for initial
times $t_i \ge -B/A$.
The condition \cite{barlid} for an inflationary phase is in
general given by (\ref{infl}). For $A>0$ we have $\dot a >0$ and find
\be
\ddot a (t) = \frac {A^2 C \exp [C \arctan (At+B)]}{(1+B^2+2ABt+A^2t^2)^2}
 \Bigl ( C - 2 A t - 2 B ) \Bigr ) \; .
\ee
In each model, depending on the constants $A,B,C$, the scale factor
starts with the value
$a(-B/A)=1$, the velocity $\dot a(-B/A)=A C$, and the acceleration
$\ddot a(-B/A)=A^2 C^2$. Then the universe inflates exponentially up
to the time $t_f=(C-2B)/(2A)$ $\Leftrightarrow \; \ddot a =0$ ($f$
means final). Then, a positive pressure of the scalar field prevents
a further expansion of the universe. Hence, the duration of inflation
is $t_f - t_i = C/(2A)$. Only the constant $C$ determines the strength of
inflation whereas both constants $A$ and $C$ determine the duration
of the inflation. The constant $B$ has no geometrical meaning.
At the end of inflation the scale factor
is constant. After infinitely long time,
in practice, very soon after the inflation phase, a Minkowski
spacetime emerges.

It is also possible to take into account a de Sitter type expansion,
i.e.~the Hubble expansion rate is becoming constant after a short
starting phase determined by the new model. We may connect the two models
by
\be
a(t) = a_0 \exp [C \arctan (A t + B) + D t + E]  \; .
\ee
Then, we find
\be
H = \frac {A C}{1+(At+B)^2} + D  \longrightarrow  D \; .
\ee
We get the same solution for the scalar field, whereas the
time--dependent potential is changed to
\ben
U(t) & = & \frac {3}{\kappa }
  \left ( \frac {A C}{1+(At+B)^2} + D \right )^2 \nonumber \\
 & & - \frac {2 A^2}{\kappa } \frac {(At + B) C}{(1+(At+B)^2)^2} \; ,
\een
so that in the limit $t \rightarrow \infty $ we have:
$U \rightarrow 3D^2/\kappa $.
The disadvantage of this model is that the de Sitter inflation very
soon plays the decisive role and inflation never ends.

\section{\bf Exact solution of new inflation}

For the graceful exit function, we consider a polynomial in $H$ up to
second order, i.e.
\be
g(H) = \frac {1}{G} H^2 + \left ( D - \frac {2A}{G} \right ) H
 + \frac {A^2}{G} - AD \; . \label{gnewh}
\ee
where $A,D,G$ are constants.
Again, it is possible to calculate the model completely.
For the Hubble expansion rate, we find
\be
H(t) = A - \frac {DG \exp (D t + F)}{1 + \exp (D t + F)} \; .
\ee
In the limit of infinitely long time, $H$ is becoming the constant $A-DG$.
For this universe the scale factor reads (Fig.~\ref{fig.2.1})
\be
a(t) = a_0
\frac {\exp (A t + K)}{\left ( 1 + \exp (D t + F) \right )^G} \; .
\label{exact}
\ee

$\!$From (\ref{phiH}) we get the scalar field (Fig.~\ref{fig.2.2})
\be
\phi (t) = \pm \sqrt { \frac {8G}{\kappa} }
  \arctan \left ( \exp \left ( \frac {D t + F}{2} \right ) \right ) + C \; .
\ee
The constants $F,K,C$ are further integration constants.
We read off the reality condition $G>0$.

The potential is given by
\ben
U(\phi ) & = & \frac {1}{\kappa \left [ 1 + \tan^2
 \left ( \pm \sqrt {\kappa } (\phi - C)/\sqrt{8G} \right ) \right ]^2}
 \Biggl [ 3 A^2  \nonumber \\
& & + \left ( 6 A^2 - 6 ADG - D^2 G \right )
 \tan^2 \left ( \pm \sqrt {\frac {\kappa }{8G}} (\phi - C) \right )
    \nonumber \\
& & + 3 \left ( A - DG \right )^2
 \tan^4 \left ( \pm \sqrt {\frac {\kappa }{8G}} (\phi - C) \right )
    \Biggr ] \; , \label{unew}
\een
cf.~Fig.~\ref{fig.2.3}.

The constants $A$ and $D$ have the dimension $1/length$.
For large times, the scale factor (\ref{exact}) reaches the value
\be
a(t) \simeq a_0
  \, \exp \biggl [ (A-DG)t \biggr ] \; . \label{ainf}
\ee
One recognizes that for $A\neq DG$ the scale factor is either
exponential in- or decreasing depending on the sign of $A-DG$.
Hence, in these cases we find models with either in- or deflationary
behavior. Only for $A=DG$, we have a limiting value. In this case,
the constant $a_0$ determines the limiting value $a_\infty $.
 From the potential (\ref{unew}), we can distinguish three
types of different physical behavior of our model universe.

Hence, in this model, a fine--tuning problem arises. Only if three
constants fulfill the exact condition $A=DG$, the graceful exit is
secured. In all other cases, the inflation can first stop, then
occurs again or the resulting universe will recollapse to its original
state.

For $A=DG$, the potential $U(\phi )$ is vanishing with increasing
scalar field. These models are the exact solutions of the
new inflationary theory.

For $A<DG$, the potential $U(\phi )$ reaches a positive value with increasing
scalar field. Therefore a {\em deflation} of the universe is born later.
The chronology of this model is:
inflation, Friedmann cosmos, deflation. The universe contracts again
after a maximal radius.

For $A>DG$, the inflationary phase first ends, then it is
renewed and never ends. In order to be precise, for increasing
deviations of $A$ from $DG$, the limiting value of the potential becomes
more and more negatively, and finally the minimum of the potential
vanishes. Hence we find the chronology of
inflation, Friedmann cosmos, inflation. The duration of the
Friedmann phase decreases with increasing $A$ and fixed product $DG$.
The slow--roll approximation occurs again.

Because all models start with an inflation phase, we can first
calculate the end of inflation, which we define by $\ddot a (t_f)=0$.
It occurs at the time $t_f$ which is implicitly given by
\be
e^{D t_f + F} = \frac {1}{2 (A-DG)^2}
\Biggl [ - 2A^2 + 2ADG + D^2G \nonumber
\ee
\be
\pm \sqrt{D^2 G (D^2 G + 4 ADG - 4A^2)} \Biggr ] \; .
\ee
For the case $A=DG$ we find
\be
t_f = - \frac {F}{D} + \frac {1}{D} \ln G  \; .
\label{tf}
\ee

There exist several constraints on the constants in order to fulfill
today's astrophysical constraints \cite{lin90}. We suppose
that inflation ends at $t_f:=\tau = 10^{-35}$s. At the Planck time
$t_i=10^{-43}$s the initial extension of the universe was the Planck
length $\ell_0 \simeq 10^{-33}$cm where the inflation has started.
Nowadays, the Hubble expansion rate is
$H(t_f) = 3.24 \times 10^{-18}$s$^{-1}$.

Let us concentrate on the graceful exit case, i.e.~$A=DG$.
There we set $G=1$. Because the inflation is arising in the range
$10^{-35}$s, we set the constants $A=D=10^{36}/$s. In order to get at
least our size of the present universe,
Eq.~(\ref{ainf}) requires $a(t_f)\simeq = a_0 =10^{28}$cm.
We find from (\ref{tf})
\be
F = \ln G - D \tau = 130.5  \; .
\ee
Using the Planck length at $t_i$ gives the condition
$K=4.54 \times 10^{-5}$.

At the origin $\phi =0$, the potential (\ref{unew}) possesses the
following  power series expansion (the inflationary part)
\be
U(\phi ) \simeq {{3\,{A^2}}\over {\kappa }} +
   \left( {{-3\,A\,D}\over 4} - {{{D^2}}\over 8} \right) \,{\phi ^2} +
   \left( {{3\,{D^2}\,\kappa }\over {64}} +
      {{A\,D\,\kappa }\over {32\,{\it G}}} +
      {{{D^2}\,\kappa }\over {48\,{\it G}}} \right) \,{\phi ^4} +
   {{{\rm O}(\phi )}^5} \; .
\ee

It is also interesting to calculate the minimum of the potential $U$.
This minimum occurs at the time $t_{min}$
\be
e^{D t_{min} + F} = \frac {6A+D}{6DG+D-6A} \; ;
\ee
observe that no minimum exists if the right hand side is negative or zero.
For the constraint $A=DG$, we can find a
minimum in every case. For this special choice of constants, we find
\be
t_{min} = \frac {1}{D} \left [ \ln \left ( 6G+1 \right ) - F
 \right ] \; ,
\ee
and a negative value of the potential at the minimum
\be
U(t_{min})_{A=DG} = - \frac {D^2 G}{4\kappa (1+3G)} \; .
\ee

After the initial submission of the present paper,
further exact inflationary solutions were
published \cite{barr}. The first solution is a special case of
(\ref{gnewh}) for $DG=2A$, i.e.,
\be
g(H) = -2 H^2/\hat A^2 + 2\hat A^2 \lambda^2
\ee
(in the notation of \cite{barr}). The solution reads then
\ben
\phi (t) & = & \hat A \ln [\tanh (\lambda t) ] \; , \\
H(t) & = & \hat A^2 \lambda \coth (2 \lambda t) \; , \\
a(t) & = & a_0 [\sinh (2 \lambda t)]^{\hat A^2/2} \; , \\
U(\phi ) & = & \hat A^2 \lambda^2 \left [ (3\hat A^2-2)
 \cosh^2 \left (\frac {\phi }{\hat A} \right ) + 2 \right ] \; .
\een
The second solution of \cite{barr} corresponds, in our notation, to
the following ansatz:
\ben
g(H) & = & 3 \hat A^{-10/3} \lambda^{-2/3} 6^{2/3} H^{8/3}
+ \hat A^{-2} (3 \hat A^2 - 9) H^{6/3} \nonumber \\
& & - \frac {3}{2} 6^{1/3} \hat A^{-2/3} \lambda^{2/3} (\hat A^2 + 1) H^{4/3}
+ \frac {6^{2/3}}{12} \hat A^{2/3} \lambda^{4/3} H^{2/3}  \; .
\een
In the notation of \cite{barr} the solution is then
\ben
\phi (t) & = & \hat A \; \mbox{csch} (\lambda t) \; , \\
H(t) & = & \frac {\hat A^2 \lambda }{6} \coth^3 (\lambda t) \; , \\
a(t) & = & a_0 [\sinh (2 \lambda t)]^{\hat A^2/2}
 \exp \left [ - \frac {\hat A^2}{12} \coth^2 (\lambda t) \right ]  \; , \\
U(\phi ) & = & \frac {\lambda^2}{12 \hat A^2} \phi^2 (\phi^2 +\hat A^2)
  \left (\frac {\phi^4}{\hat A^4} + 2 \phi^2 + \hat A^2 -6 \right ) \; .
\een

\section {\bf Deflationary universes}

One could suppose that the Hubble expansion rate has increased in the
beginning of the universe and then became constant. Having this
idea in mind, we try the ansatz
\be
g(H) = \frac {AC}{1+\tan^2 (H/C)} = AC \cos^2(H/C) \; .
\ee
which yields
\be
H = C \arctan (At+B) \; .
\ee
Then the scale factor is
\be
a(t) = a_0 \frac { \exp \left [ (Ct+BC/A) \arctan (At+B) + F \right ] }
 {\left [ 1 +(At+B)^2 \right ]^{C/(2A)} } \; ,
\ee
and the scalar field reads
\be
\phi (t) = \sqrt { -\frac {2C}{A\kappa }} \mbox{arsinh} (At+B) - D \; .
\ee
The potential is given by (Fig.~\ref{fig.3.1})
\ben
U(\phi ) & = & \frac {1}{\kappa }
\Biggl [ 3C^2 \arctan^2 \left (
 \sinh \left (\sqrt{-\frac{A\kappa }{2C}} (\phi + D) \right ) \right ) \\
 & &  + \frac {AC}{1+
 \sinh^2 \left (\sqrt{-\frac{A\kappa }{2C}} (\phi + D) \right )}
\Biggr ] \; . \label{ud1}
\een
In order to have a real scalar field solution, we have to require $C/A<0$. A
more detailed investigation of the behavior of the scale factor shows that
for negative $At+B$ we find an inflationary phase before $a(t)$
reaches a maximum. For positive $At+B$, a deflationary phase starts.

A further solution is found if one supposes that the Hubble parameter
is not becoming constant but is increasing logarithmically
\be
H =  C \ln (At+B) \; ,
\ee
which follows from the ansatz
\be
g(H) = AC \exp (-H/C) \; .
\ee
Then we find the exact solution (Fig.~\ref{fig.3.2})
\ben
a(t) & = & a_0 (At+B)^{(Ct+BC/A)} e^{-(Ct+BC/A)+F} \; , \\
\phi (t) & = & \sqrt {-\frac{8C}{A\kappa }} \sqrt {At+B \, } - D
\; , \\
U(\phi ) & = & \frac {1}{\kappa }
 \left \{ 3C^2 \left [ \ln \left (- \frac {A\kappa }{8C} (\phi +D)^2
        \right ) \right ]^2 -
 \frac {8 C^2}{\kappa } \frac {1}{(\phi +D)^2} \right \} \; .
 \label{ud2}
\een
For a real scalar field, the constraints $C/A<0$ and
$\sqrt {At+B \, }>0$ have to be fulfilled. The inflation starts at the
time $t_i=-B/A$ with
a short increasing (because of the exponential function in the scale
factor), but then decreases very rapidly to zero (because of the
function of the type $t^t$). Hence, the solution has a purely deflationary
character. However, the discrimination between inflation and deflation
is also depending on the conformal frame \cite{gasp}.

\section {\bf Density perturbations}

For a long time
one thought that the spectrum of density perturbations is described by
the scale--invariant Harrison--Zel'dovich form
\cite{guthpi,hawking,starobinsky}. But new observations by COBE
\cite{smoot} show the possibility of small deviations. The spectra of scalar
and transverse--traceless tensor perturbations \cite{lytste,lidlyt}
are given by
\ben
P^{\frac {1}{2}}_{{\cal R}} (\hat k) & = &
 \left ( \frac {H^2}{4\pi \mid H' \mid } \right ) \Biggl |_{aH=\hat k} \; , \\
P^{\frac {1}{2}}_{g} (\hat k) & = &
 \left ( \frac {H}{2\pi } \right ) \Biggl |_{aH=\hat k} \; ,
\een
where ${\cal R}$ is the perturbation in the spatial curvature,
$H'=dH/d\phi $, and $\hat k$ the wave number. The expressions on the
right hand side have to be evaluated at that comoving scale $\hat k$
which is leaving the horizon during the inflationary phase.
The results are only valid in first order slow--roll
approximation \cite{lidlyt}. The scalar and the gravitational
spectral indices in first order approximation read
\ben
n_s & := & 1 + \frac {d \ln P_{{\cal R}}}{d \ln\hat k}
= 1 - 4 \epsilon + 2 \eta \; , \\
n_g & := & \frac {d \ln P_{g}}{d \ln\hat k}
= - 2 \epsilon \; ,
\een
where the two slow--roll parameters $\epsilon$ and $\eta $ are defined
as follows:
\ben
\epsilon & := & 3 \frac {\dot \phi^2}{2U + \dot \phi^2}
 = \frac {2}{\kappa } \left ( \frac {H'}{H} \right )^2  \; , \\
\eta & := & -3 \frac {\ddot \phi }{3 H \dot \phi }
 = \frac {2}{\kappa } \frac {H''}{H}  \; .
\een
In general, they are scale dependent and have to be evaluated at the horizon.
The parameter
$\epsilon $ describes the relation between the kinetic and the total
energy, whereas $\eta $ is a measure for the relation between the
``acceleration'' of the scalar field and its ``curvature--depending velocity''.
In the slow--roll approximation $\epsilon $ and
$\eta $ are small quantities. Actually, the phase of acceleration
(the slow--roll approximation, $\ddot a>0$) is
now equivalently to the condition $\epsilon <1$. The {\em flat spectrum} of
Harrison--Zel'dovich is obtained for $n_s=1$.

By using (\ref{dotphi}), the first slow--roll parameter is given by
\be
\epsilon= - {{g}\over{H^2}} \; .
\ee
After inserting (\ref{scalar}) and (\ref{2'}), the second parameter can be
expressed as
\be
\eta = 3 -{\kappa\over 2 H}{{d\widetilde U}\over{dH}} =
-{1 \over {2H}}{{dg}\over{dH}} \; .
\ee

Thus, the condition $\eta_s=1$ for a flat spectrum of the
Harrison--Zel'dovich type converts into the relation
\be
H\,{{dg}\over{dH}}= 4g
\ee
for the graceful exit function $g$. This Euler type relation for
homogeneous functions is solved by $g=C H^4$, which corresponds to our
earlier Eq.~(\ref{n=4}), cf.~\cite{barlid}.

In the era of inflation, the scale of the universe has to
explode at least by a factor $e^{60} \simeq 10^{30}$. The number of
e--foldings between
scalar field values $\phi_1$ and $\phi_2$ is given by
\be
\ln a(\phi_1,\phi_2)
 = - \frac {\kappa }{2} \int\limits_{\phi_1}^{\phi_2}
 \frac {H(\tilde \phi )}{H'(\tilde \phi )} d \tilde \phi  \; .
\ee

In the new two models of chaotic and new inflation, the slow--roll
phase appears for the regime of small $\phi $ values. Here we
investigate only the new inflationary model because of its explicit
$\phi $ dependence. Only for $A=DG$, we find an inflationary model
for our universe which ``workes" rather well.
Hence, we can calculate (for small $\phi $ values up to order $\phi^2$)
\ben
n_s & \simeq  & 1 - \frac {1}{\kappa G}
 - \frac {3\, \phi ^2} {8\,{\kappa G^2}}  \; , \\
n_g & \simeq  & - \frac {\phi ^2} {4\,{\kappa G^2}} \; .
\een
Observe that the constant $1/G$ as well as $\phi $ occurs in second order.
Only if $G>>1$ we find a nearly scale--invariant
spectrum. For very large $G$ we have $g=0$, i.e.~the de Sitter solution.
For $\kappa G=1$ the
scalar spectral index is very small and proportional to $\phi^2$.

Of further interest is the relative contribution of the tensor and
scalar modes to the microwave background signal. The $l$th multipole
of the spherical harmonic expansion \cite{lidlyt} of the anisotropy of the
temperature is, on this scale, given by
\be
R_l (\epsilon ) := \frac {\Sigma_l^2 (tensor)}{\Sigma_l^2 (scalar)}
 \simeq  12.4 \, \epsilon
 \simeq  \frac {12.4\,{\phi^2}} {8\,{\kappa G^2}} \; .
\ee
The scalar field $\phi $ has to be evaluated at that time scale where
the corresponding $l$th multipole leaves the horizon.
This result is very similar to the one
found for the intermediate inflation (see Eq.~(21) of Ref.~\cite{barlid}).

The relation between the relative amplitude and the scalar spectral
index is given by
\be
n_s  \simeq  1 - \frac {1}{\kappa G} - \frac {3\,R_l} {12.4}  \; . \label{nsrl}
\ee
Again, we can be read off that, for $G>>1$, we
have a scale independent spectrum.

It is quite useful to compare this result with those calculated from
other inflationary models. For the intermediate inflation,
\be
n_s  \simeq  1 + \frac {n-4}{12.4} R_l
\ee
was found \cite{barlid}, whereas for the power--law inflation
\be
n_s  \simeq  1 - \frac {R_l}{6.2}
\ee
holds. In the last case, both $n_s$ and $R_l$ are scale independent.

We can now compare these $\phi $--depending results with those emerging
from the $H$--dependence. For $A=DG$ we find
\be
\epsilon = - \frac {1}{G} \left ( 1 - \frac {DG}{H} \right ) \; .
\ee
The inflation occurs for $\mid H-DG \mid <<1$. The scalar spectral
index reads
\be
n_s = 1 + \frac {2}{G} - \frac {3D}{H} \; .
\ee
Hence, in the era of inflation,
\be
n_s  \simeq  1 - \frac {1}{G}
\ee
holds.
The relation between $R_l$ and $n_s$ reads
\be
n_s = 1 - \frac {4\,R_l} {12.4} - \frac {2\kappa }{G}
     + \frac {D\kappa }{H} \; ,
\ee
which in the inflationary era reduces to (\ref{nsrl}).

\acknowledgments
We would like to thank Peter Baekler, Friedrich W.~Hehl and Yuval Ne'eman for
helpful comments. We are grateful to Yuri Obukhov for some instructive
remarks on the critical points of our system of nonlinear first order
equations and to Vitaly N.~Melnikov for drawing our attention to
Ref.~\cite{wag}.
Research support for F.E.S.~was provided by the Deutsche
Forschungsgemeinschaft, project He $528/14-1$, whereas E.W.M.~thanks
D.~Stauffer for his leave of absence from Cologne, paid by the Canada
Council.

\appendix

\section
{\bf Solutions for constant self--interaction potential}

In this Appendix, we derive all solutions for a constant
self--interaction
potential $U(\phi )=\Lambda /\kappa$ for closed, open, and
flat universes.
Then, a first integral of the Klein--Gordon equation (\ref{scalar}) is
\be
\dot \phi = \frac {C_1}{\sqrt{\kappa } a^3} \; . \label {3}
\ee
Eq.~(\ref{1}) yields
\be
\dot a^2 + k = \frac {1}{3} \Lambda a^2
+ \frac {C_1{}^2}{6 a^4}
\; , \label{4}
\ee
where we have replaced $H$ by the scale factor $a(t)$.

The general integral of Eq.~(\ref{4}) reads
\be
 t =
 \pm \sqrt{6} \int \frac {a^2 \, da}{\sqrt{ 2 \Lambda a^6 - 6k a^4
           + C_1{}^2}}  \; ,   \label{app}
\ee
for which $C_2$ is a second integration constant; cf.~\cite[p.~731]{mtw}
for the reparametrization of the time coordinate.

For $U=0$, we obtain $H=1/(3t)$ as the solution of (\ref{doth}).

The integral (\ref{app}) belongs to the {\em hyperelliptic
integrals} \cite[part 251, 6]{groeb}.
It can be transformed by means of $y:=a^2$ into an {\em elliptic integral}
\cite[part 244]{groeb}
\be
 t = \pm \frac {\sqrt{6}}{2} \int
 \frac {y\, dy}{\sqrt{ 2 \Lambda y^4 - 6 k y^3
           + C_1{}^2 y}}  \; .
\ee
The solutions depend on the special form of the zeros of the quartic
equation
\be
2 \Lambda y^4 - 6 k y^3 + C_1{}^2 y = 0 \; .
\ee

\begin{description}

\item[Case $k=\Lambda=0$:] \hfill

The solution reads
\be
a(t) = \left (\pm C_1 \sqrt{\frac {3}{2}} t
 + C_2 \right )^{1/3}
\ee
and
\be
\phi (t) = \pm \sqrt{\frac{2}{3 \kappa }}
    \ln \left (\pm C_1\sqrt{\frac{3}{2}} t + C_2 \right )
    + C_3 \; .  \label{frei}
\ee

\item[Case $k=0$ and $\Lambda \neq 0$:] \hfill

For $\Lambda >0$, the solution is
\ben
a(t) & = & \left ( \frac {C_1{}^2}{2\Lambda} \right )^{1/6}
   \sinh^{1/3}  \left (\pm \sqrt{3 \Lambda} (t+C_2) \right ) \; , \\
\phi (t) & = & \pm \sqrt{\frac{2}{3 \kappa }}
\ln \tanh \left ( \pm \frac {\sqrt{3\Lambda}}{2} (t + C_2)
\right ) + C_3  \; .
\een

For $\Lambda <0$, the hyperbolic functions are converted into the
trigonometric ones:
\ben
a(t) & = & \left ( - \frac {C_1{}^2}{2\Lambda} \right )^{1/6}
       \sin^{1/3} \left (\pm \sqrt{- 3 \Lambda} (t+C_2) \right )
\; , \\
\phi (t) & = & \pm \sqrt{\frac{2}{3 \kappa }}
\ln \tan \left ( \pm \frac {\sqrt{- 3\Lambda}}{2} (t + C_2)
\right ) + C_3  \; .
\een

\item[Case $\Lambda=0$ and $k=\pm 1$:] \hfill

\be
t = \pm  \int \frac {a^2 \, da}{ \sqrt {C_1{}^2/6 - \varepsilon a^4}}
\; ,  \label{elliptical}
\ee
where $\varepsilon =1$, for $k=+1$, and $\varepsilon =i$,
the imaginary unit, for $k=-1$. This integral is again of elliptic type.
In terms of the elliptic integrals
\be
F(\varphi ,\widetilde k) = \int\limits_0^\varphi
   \frac {d\psi}{\sqrt{1 - \widetilde k^2 \sin^2\psi \; }}
\ee
and
\be
E(\varphi ,\widetilde k) =
 \int\limits_0^\varphi \sqrt{1 - \widetilde k^2 \sin^2\psi}
 \; d\psi
\ee
of the first and second kind, respectively, we find the solution
\ben
& & t + C_2 = \pm \frac{1}{\sqrt{2}} \frac {1}{\sqrt {\varepsilon ^3}}
    \left (\frac {C_1{}^2}{6}\, \right )^{1/4}
     \Biggl [2 E \left (\varphi ,\frac{1}{\sqrt{2}} \right ) \nonumber \\
& & \qquad - F \left (\varphi ,\frac{1}{\sqrt{2}} \right ) \Biggr ]
\; ,
\een
where $a= - (1/\sqrt {\varepsilon })
\left (C_1{}^2/6 \right )^{1/4} \cos \varphi $.

For $k=+1$, the solution is also given by the integral tables of
Gr\"obner and Hofreiter \cite[part 244, 8b15c]{groeb}, whereas
for $k=-1$, we find the solution in \cite[p.~91, part 244, 8c8]{groeb}
(with $s:=r_1=s_1=s_2=-r_2= \sqrt{C_1}/(24)^{1/4}$).

\end{description}

\begin{figure}
\caption {Classification of inflationary potentials $\widetilde U(H)$.
$\widetilde U$ is measured in units of $(1/\kappa^2)$ and $H$ in units
of $(1/\kappa^{1/2})$.} \label{fig.0}
\end{figure}

\begin{figure}
\caption {Below: The scale function $a(t)$ in units of $(a_0)$
and time $t$ in units of $(1/A)$ $(C=1)$. Above: Its second time
derivative.}
\label{fig.1.1}
\end{figure}

\begin{figure}
\caption {The scalar field $\phi (t)$ in units of $(2 C/\kappa )^{1/2}$
and time $t$ in units of $(1/A)$ $(D=0)$.} \label{fig.1.2}
\end{figure}

\begin{figure}
\caption{The numerically determined potential $U(\phi )$ in units of
$(A^2C/\kappa )$ and scalar field $\phi $ in units of
$(2 C/\kappa )^{1/2}$ where we have set $C=1$.}
\label{fig.1.3}
\end{figure}

\begin{figure}
\caption{Below: The scale factor $a(t)$ in units of $(a_0)$
and time $t$ in units of $(1/A)$ for the graceful exit solution with $A=DG$
and $F=K=0$.
Above: The second time derivative of the scale factor $a(t)$ in the
same units determines
the different parts of the universe model. For $\ddot a>0$ the
universe is inflationary.} \label{fig.2.1}
\end{figure}

\begin{figure}
\caption {The scalar field $\phi (t)$ in units of $(8G/\kappa )^{1/2}$
and $t$ in units of $(1/A)$ where we have set $C=F=0$.
It monotonically increases to a non--vanishing limit.} \label{fig.2.2}
\end{figure}

\begin{figure}
\caption{The potential $U(\phi )$ in units of $(A^2/\kappa )$ and
the scalar field $\phi $ in units of $(8G/\kappa )^{1/2}$. We have set
$G=1$ and $C_1=0$. It has all features which are
demanded by the theory of the new inflation. But with one exception:
it does not possess an increasing potential wall after the
inflationary phase. Instead we have there a limiting point which prevents
the scalar field from reaching lower points on the potential. The
limiting point simulates the potential wall for large $\phi $ values.}
\label{fig.2.3}
\end{figure}

\begin{figure}
\caption{The potential $U(\phi )$ in units of $(C^2/\kappa )$ and
the scalar field $\phi $ in units of $[-2C/(A\kappa )]^{1/2}$
for $g(H)=AC\cos^2 (H/C)$. We have put $A=C=1$ and $D=0$.}
\label{fig.3.1}
\end{figure}

\begin{figure}
\caption{The potential $U(\phi )$ in units of $(C^2/\kappa )$ and
the scalar field $\phi $ in units of $[-8C/(A\kappa )]^{1/2}$
for $g(H)=AC\exp (-H/C)$. We have put $A=C=1$ and $D=0$.}
\label{fig.3.2}
\end{figure}


\begin{references}
\bibitem
{1} A.~Einstein, quoted by G.~Gamow in: {\em My World Line},
(Viking Press, New York, 1970), p.~44.

\bibitem
{mtw} C.W.~Misner, K.S.~Thorne, and J.A.~Wheeler: {\it Gravitation}
(Freeman, San Francisco 1973).

\bibitem
{3} H.~von Seeliger, {\it M\"unch.~Ber.} {\bf 26}, 373 (1896).

\bibitem
{4} C.~Neumann: {\it ``Allgemeine Untersuchung
\"uber das Newtonsche Prinzip der Fernwirkungen''} (Leipzig, 1896).

\bibitem
{5} P.~van Nieuwenhuizen, in: {\it Relativity,
Groups, and Topology II}, eds.: B.S.~DeWitt and R.~Stora
(North--Holland, Amsterdam 1984), p. 823.

\bibitem
{6} J.~A.~Schwarz, {\it Phys. Rep.} {\bf 89}, 223 (1982).

\bibitem
{7} S.~Weinberg, {\it Rev. Mod.~Phys.~} {\bf 61}, 1 (1988).

\bibitem
{8} H.~J.~Blome and W.~Priester, {\it Astrophys. Space Sci.}
{\bf 117}, 327 (1985); W.~Priester, J.~Hoell, und H.--J.~Blome,
{\it Phys. Bl.} {\bf 45}, 51 (1989).

\bibitem{kusm}
F.~V.~Kusmartsev, E.~W.~Mielke, and F.~E.~Schunck, {\it Phys.~Rev.}
{\bf D43} 3895 (1991); {\em Phys.~Lett.} {\bf A157} 465 (1991).

\bibitem {baek}
P.~Baekler, E.~W.~Mielke, R.~Hecht, and F.~W.~Hehl, {\em Nucl.~Phys.}
{\bf B288}, 800 (1987).

\bibitem {sch2}
F.~E.~Schunck, {\em Ann.~Physik (Leipzig)} {\bf 2}, 647 (1993).

\bibitem {mie86}
E.~W.~Mielke, {\em Z.~Naturforsch.} {\bf 41a}, 777 (1986).

\bibitem {linde}
A.~Linde: {\em Inflation and Quantum Cosmology} (Academic Press, San
Diego, 1990).

\bibitem {brandenberger}
R.~H.~Brandenberger, {\em Rev.~Mod.~Phys.} {\bf 57}, 1 (1985);
in: {\em Proceedings of the 7th Swicca Summer School
in Particles and Fields}, eds.~O.~Eboli and V.~Ribelles (World
Scientific Press, Singapore, 1993); preprint BROWN--HET--893.

\bibitem {guth}
A.~H.~Guth, {\em Phys.~Rev.} {\bf D23}, 347 (1981);
in: {\em Proceedings
of the National Academy of Sciences --- Colloqium on Physical
Cosmology}, Irvine, California, 27--28 March 1992, David N.~Schramm, ed.
Vol.~90, 1993, p. 4871--4877.

\bibitem {linde82}
A.~Linde, {\em Phys.~Lett.} {\bf B108}, 353 (1982);
{\em Phys.~Lett.} {\bf B249}, 18 (1990).

\bibitem {lucmat}
F.~Lucchin and S.~Matarrese, {\em Phys.~Rev.} {\bf D32}, 1316 (1985).

\bibitem {barrow87}
J.~D.~Barrow, {\em Phys.~Lett.} {\bf B187}, 12 (1987).

\bibitem {barrow90}
J.~D.~Barrow, {\em Phys.~Lett.} {\bf B235}, 40 (1990).

\bibitem {stein}
P.~J.~Steinhardt, {\em Class.~Quantum Grav.} {\bf 10}, S33 (1993).

\bibitem {laste}
D.~La and P.~J.~Steinhardt, {\em Phys.~Rev.~Lett.} {\bf 62}, 376
(1989).

\bibitem {barmae}
J.~D.~Barrow and K.~Maeda, {\em Nucl.~Phys.} {\bf 341}, 294 (1990).

\bibitem {wald}
R.~M.~Wald, {\em Phys.~Rev.} {\bf D28}, 2118 (1983).

\bibitem {turner}
M.~S.~Turner and L.~M.~Widrow, {\em Phys.~Rev.~Lett.} {\bf 57}, 2237
(1986).

\bibitem {barrow93}
J.~D.~Barrow, {\em Phys.~Rev.} {\bf D48}, 1585 (1993).

\bibitem {lin94}
A.~D.~Linde, Phys. Rev. {\bf D49} (15. Jan. 1994).

\bibitem {cope94}
E.~J.~Copeland, A.~R.~Liddle, D.~H.~Lyth, E.~D.~Stewart, and D.~Wands,
{\em False vacuum inflation with Einstein gravity}, preprint
astro--ph/9401011.

\bibitem {mie77}
E.~W.~Mielke, {\em Fortschr.~Phys.} {\bf 25}, 401 (1977).

\bibitem{brans}
C.~Brans and R.~H.~Dicke, {\em Phys.~Rev.} {\bf 124} 925 (1961).

\bibitem {wag}
R.~V.~Wagoner, {\em Phys.~Rev.} {\bf D1}, 3209 (1970).

\bibitem{beken}
J.~D.~Bekenstein, {\em Ann.~Phys.~(N.Y.)} {\bf 82} 535 (1974).

\bibitem {kasp}
U.~Kasper and H.--J.~Schmidt, {\em Nuovo Cim.} {\bf 104B}, 563 (1989);
H.--J.~Schmidt, {\em Astron.~Nachr.} {\bf 311}, 99 (1990).

\bibitem {galt}
D.~V.~Gal'tsov and  B.~C.~Xanthopoulos, {\em J.~Math.~Phys.} {\bf 33},
273 (1992).

\bibitem {piran}
T.~Piran and R.~M.~Williams, {\em Phys.~Lett.} {\bf B163}, 331 (1985).

\bibitem {hehl}
F.~W.~Hehl, J.~D.~McCrea, E.~W.~Mielke, and Y.~Ne'eman:
``Metric--affine gauge theories of gravity: Field equations, Noether
identities, world spinors, and breaking of dilation invariance'',
Physics Reports submitted (1993).

\bibitem {obuk}
Yu.~N.~Obukhov, {\em Phys.~Lett.} {\bf A182}, 214 (1993).

\bibitem {kolb}
E.~J.~Copeland, E.~W.~Kolb, A.~R.~Liddle, and J.~E.~Lidsey,
{\em Phys.~Rev.} {\bf D48}, 2529 (1993).

\bibitem {lid}
J.~E.~Lidsey, {\em Phys.~Lett.} {\bf B273}, 42 (1991).

\bibitem {gott}
S.~Gottl\"ober, {\em Ann.~Physik (Leipzig)} {\bf 45}, 452 (1988).

\bibitem {lin83}
A.~D.~Linde, {\em Phys.~Lett.} {\bf B129}, 177 (1983).

\bibitem {steph}
H.~Stephani: {\em General Relativity --- An Introduction to the theory
of the gravitational field} (Cambridge University Press, Cambridge,
1982).

\bibitem {barlid}
J.~D.~Barrow and A.~R.~Liddle, {\em Phys.~Rev.} {\bf D47}, 5219 (1993).

\bibitem {lin90}
A.~D.~Linde: {\it Particle Physics and Inflationary Cosmology}
(Harwood Academic Publishers, Chur, Switzerland 1990).

\bibitem {barr}
J.D.~Barrow, {\em Phys.~Rev.} {\bf D49}, 3055 (1994).

\bibitem {gasp}
M.~Gasperini and G.~Veneziano, {\em Mod.~Phys.~Lett.} {\bf A8}, 3701
(1993).

\bibitem {guthpi}
A.~H.~Guth and S.--Y.~Pi, {\em Phys.~Rev.~Lett.} {\bf 49}, 1110
(1982).

\bibitem {hawking}
S.~W.~Hawking, {\em Phys.~Lett.} {\bf B115}, 339 (1982).

\bibitem {starobinsky}
A.~A.~Starobinsky, {\em Phys.~Lett.} {\bf B117}, 175 (1982);
{\em Sov.~Astron.~Lett.} {\bf 11}, 323 (1985).

\bibitem {smoot}
G.~F.~Smoot et al., {\em Astr.~J.~Lett.} {\bf 396}, L1 (1992).

\bibitem {lytste}
D.~H.~Lyth and E.~D.~Stewart, {\em Phys.~Lett.} {\bf B274}, 168 (1992).

\bibitem {lidlyt}
A.~R.~Liddle and D.~H.~Lyth, {\em Phys.~Lett.} {\bf B291}, 391 (1992).

\bibitem {groeb}
W.~Gr\"obner and N.~Hofreiter: {\em Integraltafel} (Springer--Verlag,
Wien, 1965).


\end{references}
\end{document}